\documentclass{article}

\usepackage{microtype}
\usepackage{graphicx}
\usepackage{subfigure}
\usepackage{booktabs} % for professional tables
\usepackage{multirow}
\usepackage{arydshln}

\usepackage{microtype}
\usepackage{graphicx}
\usepackage{subfigure}
\usepackage{booktabs} % for professional tables

\usepackage[inline]{enumitem}
\usepackage{hyperref}

\usepackage[accepted]{icml2022}

\usepackage{amsmath}
\usepackage{amssymb}
\usepackage{mathtools}
\usepackage{amsthm}

\usepackage[capitalize,noabbrev]{cleveref}

\theoremstyle{plain}

\theoremstyle{definition}

\theoremstyle{remark}

\usepackage[accepted]{icml2022}

\icmltitlerunning{Dynamic Restrained Uncertainty Weighting Loss for Multitask Learning of Vocal Expression}

\begin{document}

\twocolumn[
\icmltitle{Dynamic Restrained Uncertainty Weighting Loss \\ for Multitask Learning of Vocal Expression}

\begin{icmlauthorlist}
\icmlauthor{Meishu Song}{yyy,zzz}
\icmlauthor{Zijiang Yang}{yyy}
\icmlauthor{Andreas Triantafyllopoulos}{yyy}
\icmlauthor{Xin Jing}{yyy}
\icmlauthor{Vincent Karas}{yyy}
\icmlauthor{Xie Jiangjian}{yyy,aaa}
\icmlauthor{Zixing Zhang}{xxx}
\icmlauthor{Yamamoto Yoshiharu}{yyy}
\icmlauthor{Bj\"{o}rn W. Schuller}{yyy,xxx}

\end{icmlauthorlist}

\icmlaffiliation{yyy}{EIHW, University of Augsburg, Augsburg, Germany}
\icmlaffiliation{zzz}{Educational Physiology Laboratory, University of Tokyo, Japan}
\icmlaffiliation{xxx}{GLAM, Imperial College, London, United Kingdom}
\icmlaffiliation{aaa}{School of Technology, Beijing Forestry University, Beijing, China}

\icmlcorrespondingauthor{JJ.X}{shyneforce@bifu.edu.cn, meishu.song@uni-a.de}
\icmlkeywords{multitask learning, CNN}

\vskip 0.3in
]

\printAffiliationsAndNotice{\icmlEqualContribution} 

\begin{abstract}

We propose a novel Dynamic Restrained Uncertainty Weighting Loss to experimentally handle the problem of balancing the contributions of multiple tasks on the ICML ExVo 2022 Challenge.
The multitask aims to recognise expressed emotions and demographic traits from vocal bursts jointly. 
Our strategy combines the advantages of Uncertainty Weight and Dynamic Weight Average, by extending weights with a restraint term to make the learning process more explainable. 
We use a lightweight multi-exit CNN architecture to implement our proposed loss approach. The experimental H-Mean score (0.394) shows a substantial improvement over the baseline H-Mean score (0.335).

\end{abstract}

\section{Introduction}
\label{Multitask review--advantages}
 ``Transfer should always be useful"; any pair of distributions underlying a pair of tasks must have \texttt{something} in common~\cite{mahmud2009universal}. Many deep learning works attempt to improve their performance using Multitask Learning (MTL) -- a form of learning which constitutes the learning of several related tasks simultaneously. MTL has been applied successfully in a variety of fields, including emotion recognition \cite{Eyben2012}, \cite{Chen2017}, \cite{Deng2020}, \cite{Shen2021}, visual scene understanding \cite{Liu2019b}, \cite{kendall2018multi} and automatic speech recognition \cite{Krishna2018}. 
The multitasking learning strategy has several advantages compared with single task learning. It allows the model to discover general representations that are useful for multiple tasks, giving it greater contextual knowledge and improving performance in the individual tasks compared to single-task learning \cite{kendall2018multi}. In addition, the shared parts of the network architecture make this approach more computationally efficient than training a separate model for each task, which is important in settings where the model has to run in real time \cite{kendall2018multi}.

Even though multitasking has demonstrable theoretical advantages, there are two main issues in applying multitasking to real works \cite{Liu2019b}. 
  1. Define what to share. Traditional MTL algorithms usually assume that all the tasks are related, which means tasks are sharing the same embeddings. However, the same low-level features or high-level features cannot always perform well for all tasks.
  2. Define how to balance tasks. Prior MTL works treated all tasks with equal weights, which leads to easier tasks overfitting while tough tasks are still training. 
  In other words, model performance is extremely sensitive to weight selection for balancing the individual losses.

Various prior works have attempted to address these issues.
One existing approach is using multi-exit architectures for different objectives~\cite{phuong2019distillation}, in which a stack of processing output layers is interleaved with early output layers.
The work \cite{Liu2019b} introduced a CNN-based framework that involves multiple levels of a shared network, in order to learn a global feature set and the contributions of features with varying complexity for each specific task. Both of these methods can help determine the most appropriate output layer for different tasks. In this regard, we propose a multi-exit CNN to solve the first issue.

Loss weighting strategies in MTL can be classified based on task gradients and loss: (1) pure loss strategies including Uncertainty Weighting (UW) \cite{kendall2018multi} and  Dynamic Weight Average (DWA) \cite{Liu2019b}, and (2) loss based on gradients including Gradient Normalisation \cite{chen2018gradnorm}, Multiple-Gradient Descent Algorithm \cite{sener2018multi}, Projecting Conflicting Gradient \cite{yu2020gradient}, Gradient Sign Dropout \cite{chen2020just}, Impartial Multi-Task Learning \cite{liu2021towards}, and Gradient Vaccine \cite{wang2020gradient}.
In this work, we limit our analysis to pure loss strategies since conflicting gradient signals coming from different tasks can degrade model performance~\cite{gong2019comparison}. In comparison, loss strategies, such as UW and DWA, are easy to be implemented, and continue to be popular in the literature \cite{gong2019comparison}.
We aim to combine the advantages of UW and DWA by extending UW with a term restraining the loss weights, which improves performance while also making the learning process more explainable.  

 Our contributions are two-fold:  We use a multi-exit CNN to learn data representations at multiple layers, resulting in a simple, effective, and lightweight model that achieves competitive performance. We propose a novel loss weighting strategy, namely the Dynamic Restrained Uncertainty Weighting loss for balancing multiple tasks, which incorporates both UW's and DWA's benefits.

\section{Dataset: ExVo Multitask Challenge}\label{sec:dataset}

The dataset\footnote{We use the original release of processed .wav files not the follow-up release of unprocessed ones.} used for the ExVo Multitask challenge is the HUME-VB competition dataset~\cite{baird2022icml}, a large-scale dataset of emotional human vocal bursts, which was collected in 
%BS: unified to BE...
four countries with broadly different cultures: China, South Africa, USA, and Venezuela. There are 59,299 recordings totalling 36:50:40 (HH:MM:SS) of audio data from 1702 speakers, aged from 20 to 39 years old. Furthermore, the data was recorded in speakers' homes with their own microphones. Each vocal burst had been self-annotated by each speaker in terms of the intensity of ten different expressed emotions, each on a 1-100 scale, Amusement, Awe, Awkwardness, Distress, Excitement, Fear, Horror, Sadness, Surprise, and Triumph.
In this paper, we utilise this dataset to train a multi-task model to jointly predict the expression of ten emotions along with the age and native-country of the speaker.

\section{Methodology}\label{sec:methodology}
We now introduce our multi-exit CNN for MTL, which addresses the ``what to share" issue and Dynamic Restraint Uncertainty Weighting (DRUW) loss to address the ``how to balance task weights" issue.

\subsection{Multi-Exit CNN Architecture}

\begin{figure}[t]
    \centering
    \includegraphics[width=0.47\textwidth]{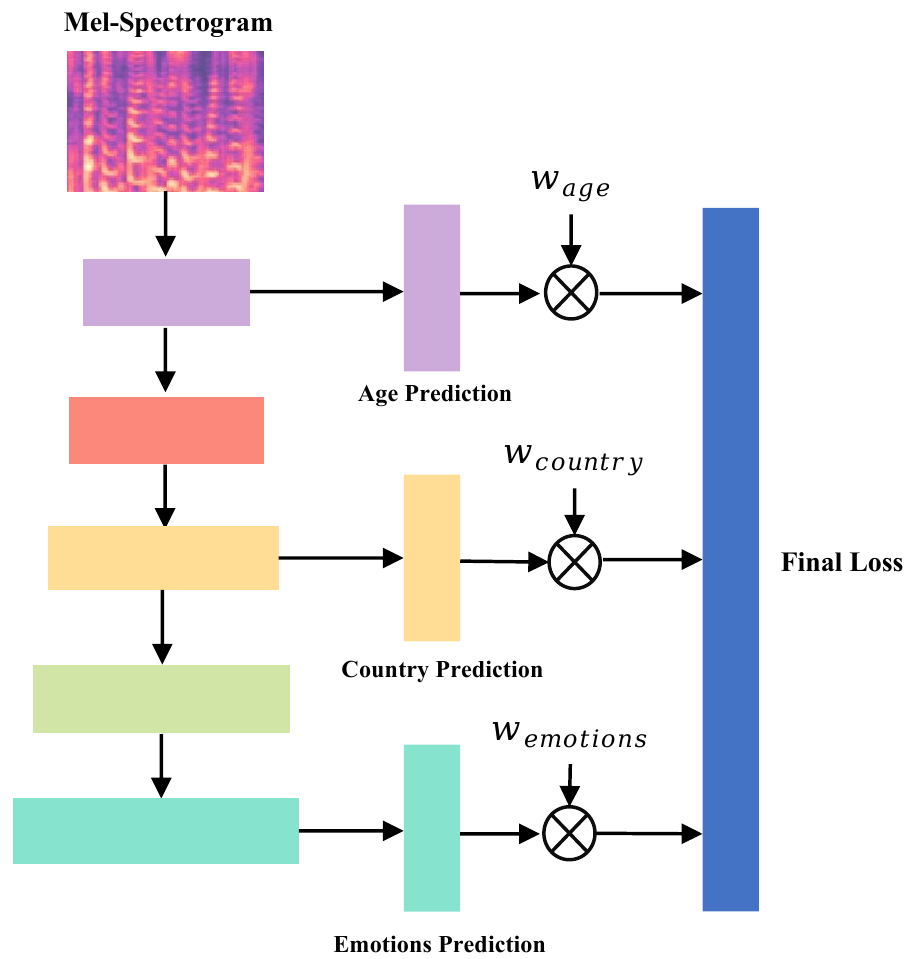}
    \caption{Overview of the multi-exit CNN architecture.}
    \label{fig:arch}
\end{figure}

CNN architectures are naturally hierarchical~\cite{zhu2017b}. 
In~\cite{zeiler2014visualizing}, it is shown that lower layers in CNNs usually capture the low-level features, while higher layers are likely to extract high-level features. As a consequence, a possible way to embed different tasks into a single CNN model is to obtain predictions from different CNN layers as data flow through it~\cite{zhu2017b}, thus effectively transforming a single, monolithic architecture to a scalable, multi-exit one. 
Inspired by the flexibility of controlling output layers of a CNN, we propose a multi-exit CNN model by adding different output layers to learn different data representations. The overview of the architecture is shown in ~\cref{fig:arch}. 
Our network is composed of five hierarchical CNN blocks which contains a total of ten CNN layers, and each block has two layers. The Age prediction is branched off after the first Conv block, the Country prediction after the third Conv block, while the emotion prediction after the fifth Conv block. 
One important aspect of this work is determining the depth of the exit branch for each task. This is done experimentally by selecting the depth which gives the best performance.
Due to the layer number being relatively low, we apply a grid search technique based on the validation set performance to find the proper exits for each task.

\subsection{Loss Weighting Strategies}

% Our proposed DRUW loss is a reworked version of two extensively adopted loss strategies: UW and DWA losses.
% In this section, we first introduce these two strategies and then introduce our loss strategy. 

\subsubsection{Uncertainty Weighting Loss}

The performance of MTL models is largely determined by shared weights, according to~\cite{kendall2018multi}, but these weights are difficult to train.
This leads to the concept of uncertainty to measure the loss of different tasks, making it possible to learn different types of tasks simultaneously~\cite{kendall2018multi}. UW is defined as in \cref{eq:UW}:

\begin{equation}
\begin{aligned}
\mathcal{L}(w,\alpha_{1},\alpha_{2}) &= \frac{1}{\alpha_{1}^{2}}\mathcal{L}_{1}(w) + \frac{1}{\alpha_{2}^{2}}\mathcal{L}_{2}(w) \\
&\quad + log\alpha_{1} + log\alpha_{2}.
\end{aligned}
\label{eq:UW}
\end{equation}

The authors interpret $ \alpha_{1}$ and $\alpha_{2} $
as learning the relative weight of the losses $\mathcal{L}_{1}(w)$ and $\mathcal{L}_{2}(w)$ adaptively ~\cite{kendall2018multi}, based on the data. 
For a task with a larger loss and higher uncertainty, the strategy can effectively avoid the model to ``take a big step" towards a ``blind" direction. Thus, it should update $w$ with a smaller gradient. At the same time, the strategy can avoid the problem of a task with larger loss dominating the overall loss.
Recently, the Revised Uncertainty Weighting (RUW) Loss~\cite{liebel2018auxiliary} was proposed to solve a major disadvantage of the UW loss: experimentally, because the log part in \cref{eq:UW} leads the whole loss to a negative value~\cite{liebel2018auxiliary}, the RUW adapts the regularisation term $log\alpha$ to $log(1+log\alpha^{2})$ in order to enforce positive values. The concept can be represented as in \cref{eq:RUW}:

\begin{equation}
\begin{aligned}
\mathcal{L}(w,\alpha_{1},\alpha_{2}) &= \frac{1}{\alpha_{1}^{2}}\mathcal{L}_{1}(w) + \frac{1}{\alpha_{2}^{2}}\mathcal{L}_{2}(w) \\
&\quad + log(1+log\alpha_{1}^{2}) + log(1+log\alpha_{2}^{2}).
\end{aligned}
\label{eq:RUW}
\end{equation}

\subsubsection{Dynamic Weight Average}
In~\cite{Liu2019b}, Dynamic Weight Average (DWA) is proposed as a simple, yet effective
adaptive weighting method. Inspired by GradNorm~\cite{chen2018gradnorm}, this learns to average task weighting over time by evaluating the rate of
change of individual loss.
In detail, a weighting $\lambda_{k}$ is defined for each task $k$ as:
\begin{equation}
\lambda_{k}(t) = \frac{\mathcal{K} exp(\frac{\mathcal{L}_{k}(t-1)}{\mathcal{L}_{k}(t-2)}/\mathcal{T})}{\sum_{i}exp(\frac{\mathcal{L}_{k}(t-1)}{\mathcal{L}_{k}(t-2)}/\mathcal{T})},
\end{equation}
where $t$ is an iteration index and $w_{k}$ represents the relative descending rate in the range (0, +$\infty$). 
First, 
$\mathcal T$ plays the role of smoothing task weights; the larger $\mathcal T$ is, the more evenly distributed the weights of different tasks are. Finally, the softmax operator multiplies by $\mathcal{K}$ to ensure that  $\sum_{i}\lambda_{i}(t) = K. $
In this setting, the total loss value is the sum of DWA weights multiply each task loss.

\begin{table*}
\centering
\caption{ExVo-MultiTask task validation and test results.
Mean concordance correlation coefficient (CCC) across the ten (Emo)tional classes, unweighted average recall (UAR) for the four class (Cou)try task (chance level 0.25), and mean absolute error (MAE) for the Age regression task are all reported. H-Mean score means harmonic mean between these metrics. Emphasised results indicate best scores. Besides official baselines, we report Single Task (ST) baselines and MTL baselines using our proposed multi-exit CNN model. For comparison, we report weighting strategies: Equal Weighting (EW), Uncertainty Weighting (UW), Revised Uncertainty Weighting (RUW), our Restrained Uncertainty Weighting (RRUW), Dynamic Weight Average (DWA), and our Dynamic Restrained Uncertainty Weighting (DRUW).}

\begin{tabular}{llcccccc} 
\hline\hline
                           &                  & \multicolumn{5}{c}{\textbf{Validation}}                              & \multicolumn{1}{c}{\textbf{Test}}  \\
                           & Weighting~       & Emo-CCC        & Cou-UAR        & Age-MAE        & H-Mean         &  & H-Mean                             \\ 
\hline
Official Baselines ComPARE & EW               & 0.416          & 0.506          & 4.222          & 0.349          &  & 0.335                              \\
Official Baselines eGeMAPS & EW               & 0.353          & 0.423          & 4.011          & 0.324          &  & 0.314                              \\
Our CNN ST Baselines          & ---               & \textbf{0.645} & \textbf{0.588} & 3.926          & 0.418          &  & 0.393                              \\
Our MTL Baselines          & EW               & 0.633          & 0.525          & 3.928          & 0.405          &  & 0.382                              \\ 
\cline{2-8}
\multirow{5}{*}{Our MTL}   & UW               & 0.615          & 0.575          & 4.024          & 0.406          &  & 0.391                              \\
                           & RUW              & 0.629          & 0.539          & 3.798          & 0.414          &  & 0.391                              \\
                           & \textbf{Our RRUW}      & 0.635          & 0.576          & 3.803          & 0.421          &  & 0.392                              \\
                           & DWA              & 0.637          & 0.545          & 3.754          & 0.419          &  & 0.389                              \\
                           & \textbf{Our DRUW} & 0.635          & 0.570          & \textbf{3.763} & \textbf{0.423} &  & \textbf{0.394}                     \\
\hline\hline
\end{tabular}
\label{ref:tab}
\end{table*}

\subsubsection{Proposed Loss Strategy}

We firstly use all previously described approaches for our task. We notice that both, the UW and RUW methods, lead to a similar outcome: since the weights are also trainable, if all weights turn to decrease simultaneously, the whole loss value is also decreasing, thus leading to a trivial solution.
Thus, we propose to extend the UW and RUW with a constraint. 
Similarly with DWA, it uses a  ``softmax" output that converts the logits of the loss change ratio into a weight for each task.
Also, in the original paper of UW~\cite{kendall2018multi}, the UW formula can be ``scaled up". In more detail, using softmax, all losses can be summed up to one. ``Scaled" is a term used to describe a situation 
%BS: added (did also many smaller changes above w/o noting)
where 
all losses can be added up to a customised positive value. Differently, this is suggested in a classification task~\cite{kendall2018multi}. In our case, the task is a heterogeneous multitask problem which involves regression and classification tasks in parallel.
To constrain the weights in a controlled manner, we use a different strategy with softmax, and we simply add the following $\varphi$ with RUW, namely Restrained Revised Uncertainty Weighting (RRUW) Loss:

\begin{equation}
\begin{aligned}
\mathcal{L}(w,\alpha_{1},\alpha_{2}) &= \frac{1}{\alpha_{1}^{2}}\mathcal{L}_{1}(w) + \frac{1}{\alpha_{2}^{2}}\mathcal{L}_{2}(w) \\
&\quad + log(1+log\alpha_{1}^{2}) + log(1+log\alpha_{2}^{2}) \\ 
&\quad + \left |  \varphi - (\left | log\alpha_{1} \right | + \left | log\alpha_{2} \right |  ) \right |.
\end{aligned}
\label{eq:loss_sum}
\end{equation}

Our method regularises the weights by guiding their sum towards a fixed positive value $\varphi$, which can constrain the sum of all weights so that we avoid the problem of the weights degenerating to a trivial solution.
To maintain a positive final loss, we additionally use the absolute value of $\varphi$ for the loss. 
Unlike softmax, which adds up all weights into one, our method controls the weight sum into a `flexible' positive value of $\varphi$. 
The term `flexible' refers to the fact that the sum of weights can be indefinitely near to $\varphi$, the fixed value.
Experimentally, DWA is shown to effectively decrease over-fitting effects of overly complex networks for easy tasks, while RRUW can uncover good data representations for complex tasks. This leads to the question: ``how can we best combine both strategies?''. We extend our idea to get sum values of RRUW and DWA weights for each task, which means for each training, there are two weight suggestion values. The joint loss is given as: 
\begin{equation}
\begin{aligned}
\mathcal{L}(w,\alpha_{1},\alpha_{2})&= 
\frac{\mathcal{K} exp(\frac{\mathcal{L}_{1}(t-1)}{\mathcal{L}_{1}(t-2)}/\mathcal{T})}{\sum_{i}exp(\frac{\mathcal{L}_{1}(t-1)}{\mathcal{L}_{1}(t-2)}/\mathcal{T})}\mathcal{L}_{1}+\frac{1}{\alpha_{1}^{2}}\mathcal{L}_{1}(w) \\ 
&\quad + \frac{\mathcal{K} exp(\frac{\mathcal{L}_{2}(t-1)}{\mathcal{L}_{2}(t-2)}/\mathcal{T})}{\sum_{i}exp(\frac{\mathcal{L}_{2}(t-1)}{\mathcal{L}_{2}(t-2)}/\mathcal{T})}\mathcal{L}_{2}+\frac{1}{\alpha_{2}^{2}}\mathcal{L}_{2}(w)\\ 
&\quad + log(1+log\alpha_{1}^{2}) + log(1+log\alpha_{2}^{2}) \\ 
&\quad + \left |  \varphi - (\left | log\alpha_{1} \right | + \left | log\alpha_{2} \right |  ) \right |.
\end{aligned}
\end{equation}

\subsection{Experimental Set-Up}
We extracted Mel-Spectrogram features by randomly cutting raw audios into 2.5 seconds length. The size of the Mel-Spectrogram is: 64*512. In the training session, we use AdamW as our optimiser. The initial learning rate is 0.001; the batch size is 32. All of our models are trained on Nvidia RTX3090 and Nvidia A40 GPUs, and the number of training epochs are limited to 60 on all tasks. 
In loss setting, the $\varphi$ value is 1 and $\mathcal T$ is 10.
In our implementation, all models are trained without data augmentation and no ensemble methods are adopted. 
\section{Results}\label{sec:results}

For the baseline results~\cite{baird2022icml} in Table~\ref{ref:tab}, we present four types including two official baselines -- Single Task (ST) and MTL baselines --, along with our proposed multi-exit CNN networks. 
On single Emotions and the Country task (CCC: 0.645 and UAR: 0.588), they outperform MTL (CCC: 0.635 and UAR: 0.570), supported by previous work~\cite{zhang2019attention}, MTL does not always guarantee a better performance. However, on the Age task, MTL (MAE: 0.3763) is superior to the one on ST (MAE: 3.928). Furthermore, MTL (Validation: 0.423, Test: 0.394) presents better H-Mean scores compared with ST (Validation: 0.418, Test: 0.393). Overall, our results confirm the advantage of MTL over ST -- it can build well-shared data representations with limited training resources~\cite{gong2019comparison}.

We also present MTL results with various loss weighting schemes. First, comparing to EW, all the strategies show improvements on H-Mean scores. This suggests that the two main stream losses, as well as our proposed losses, can all be effective in balancing task weights during training.
Second,  it is noted that our proposed multi-exit CNN with RRUW weighting strategy achieves the better H-Mean scores (Validation: 0.421 and Test: 0.392) than the previous strategies (UW, RUW, and DWA).
A rational behind this is that our ``Restraint" component can assist the model to balance the weight in a regulated manner and provide easily explainable features of the model.
In addition, we observe that, when integrating with the DRUW (Validation H-Mean: 0.423 and Test H-Mean: 0.394)
%BS: please add the UNIT (CCC?)
strategy, the system performance results in better H-Mean scores than the RRUW strategy (Validation H-Mean: 0.421 and Test H-Mean: 0.392). This indicates that as we expected, the DRUW method can keep task weights well balanced and combine the advantages of UW and DWA, as well as keep UW in a controlled way due to the ``Restraint" component and generate an easily explainable data representation. 

\section{Conclusion}\label{sec:conclusion}

In this work, we contributed a novel loss weighting strategy: DRUW, which applies on a light and effective multi-exit CNN network. 
Our proposed method combines the previous loss strategies: UW and DWA, and also gained a restrained feature for a clear explainable purpose.
The experimental results demonstrate a considerable improvement over the baseline results, which indicates the effectiveness of our methods on the ExVo Multitask challenge.
Future work includes automatic adjustment of CNN multi exits on each task. In addition, we also intend to explore the comparison between gradient oriented loss strategies and our loss strategies. Furthermore, the automatic balancing of the weights of RRUW and DWA is also promising in the broader MTL field. 
%BS: some parts needed a lot of language correction - please mark any further changes from now on :) Great work - thanks!

\pagebreak
\bibliography{example_paper}

\begin{thebibliography}{21}
\providecommand{\natexlab}[1]{#1}
\providecommand{\url}[1]{\texttt{#1}}
\expandafter\ifx\csname urlstyle\endcsname\relax
  \providecommand{\doi}[1]{doi: #1}\else
  \providecommand{\doi}{doi: \begingroup \urlstyle{rm}\Url}\fi

\bibitem[Baird et~al.(2022)Baird, Tzirakis, Gidel, Jiralerspong, Muller,
  Mathewson, Schuller, Cambria, Keltner, and Cowen]{baird2022icml}
Baird, A., Tzirakis, P., Gidel, G., Jiralerspong, M., Muller, E.~B., Mathewson,
  K., Schuller, B., Cambria, E., Keltner, D., and Cowen, A.
\newblock The icml 2022 expressive vocalizations workshop and competition:
  Recognizing, generating, and personalizing vocal bursts.
\newblock \emph{arXiv preprint arXiv:2205.01780}, 2022.

\bibitem[Chen et~al.(2017)Chen, Jin, Zhao, and Wang]{Chen2017}
Chen, S., Jin, Q., Zhao, J., and Wang, S.
\newblock Multimodal multi-task learning for dimensional and continuous emotion
  recognition.
\newblock In \emph{Proc. Annual Workshop on Audio/Visual Emotion Challenge},
  pp.\  19–26, New York, NY, USA, 2017.

\bibitem[Chen et~al.(2018)Chen, Badrinarayanan, Lee, and
  Rabinovich]{chen2018gradnorm}
Chen, Z., Badrinarayanan, V., Lee, C.-Y., and Rabinovich, A.
\newblock Gradnorm: Gradient normalization for adaptive loss balancing in deep
  multitask networks.
\newblock In \emph{Proc. International Conference on Machine Learning}, pp.\
  794--803, Stockholm, Sweden, 2018.

\bibitem[Chen et~al.(2020)Chen, Ngiam, Huang, Luong, Kretzschmar, Chai, and
  Anguelov]{chen2020just}
Chen, Z., Ngiam, J., Huang, Y., Luong, T., Kretzschmar, H., Chai, Y., and
  Anguelov, D.
\newblock Just pick a sign: Optimizing deep multitask models with gradient sign
  dropout.
\newblock \emph{Advances in Neural Information Processing Systems},
  33:\penalty0 2039--2050, 2020.

\bibitem[Deng et~al.(2020)Deng, Chen, and Shi]{Deng2020}
Deng, D., Chen, Z., and Shi, B.~E.
\newblock Multitask emotion recognition with incomplete labels.
\newblock In \emph{Proc. Automatic Face and Gesture Recognition}, pp.\
  592--599, Virtual, 2020.

\bibitem[Eyben et~al.(2012)Eyben, W\"{o}llmer, and Schuller]{Eyben2012}
Eyben, F., W\"{o}llmer, M., and Schuller, B.
\newblock A multitask approach to continuous five-dimensional affect sensing in
  natural speech.
\newblock \emph{ACM Trans. Interact. Intell. Syst.}, 2\penalty0 (1), March
  2012.
\newblock ISSN 2160-6455.

\bibitem[Gong et~al.(2019)Gong, Lee, Stephenson, Renduchintala, Padhy,
  Ndirango, Keskin, and Elibol]{gong2019comparison}
Gong, T., Lee, T., Stephenson, C., Renduchintala, V., Padhy, S., Ndirango, A.,
  Keskin, G., and Elibol, O.~H.
\newblock A comparison of loss weighting strategies for multi task learning in
  deep neural networks.
\newblock \emph{IEEE Access}, 7:\penalty0 141627--141632, 2019.

\bibitem[Kendall et~al.(2018)Kendall, Gal, and Cipolla]{kendall2018multi}
Kendall, A., Gal, Y., and Cipolla, R.
\newblock Multi-task learning using uncertainty to weigh losses for scene
  geometry and semantics.
\newblock In \emph{Proc. Computer Vision and Pattern Recognition}, pp.\
  7482--7491, Salt Lake City, UT, USA, 2018.

\bibitem[Krishna et~al.(2018)Krishna, Toshniwal, and Livescu]{Krishna2018}
Krishna, K., Toshniwal, S., and Livescu, K.
\newblock Hierarchical multitask learning for ctc-based speech recognition.
\newblock \emph{CoRR}, abs/1807.06234, 2018.
\newblock URL \url{http://arxiv.org/abs/1807.06234}.

\bibitem[Liebel \& K{\"o}rner(2018)Liebel and K{\"o}rner]{liebel2018auxiliary}
Liebel, L. and K{\"o}rner, M.
\newblock Auxiliary tasks in multi-task learning.
\newblock \emph{arXiv preprint arXiv:1805.06334}, 2018.

\bibitem[Liu et~al.(2021)Liu, Li, Kuang, Xue, Chen, Yang, Liao, and
  Zhang]{liu2021towards}
Liu, L., Li, Y., Kuang, Z., Xue, J., Chen, Y., Yang, W., Liao, Q., and Zhang,
  W.
\newblock Towards impartial multi-task learning.
\newblock In \emph{Proc. International Conference on Learning Representations},
  Vienna, Austria, 2021.

\bibitem[Liu et~al.(2019)Liu, Johns, and Davison]{Liu2019b}
Liu, S., Johns, E., and Davison, A.~J.
\newblock End-to-end multi-task learning with attention.
\newblock In \emph{Proc. Computer Vision and Pattern Recognition}, Long Beach,
  CA, USA, 2019.

\bibitem[Mahmud(2009)]{mahmud2009universal}
Mahmud, M.~H.
\newblock On universal transfer learning.
\newblock \emph{Theoretical Computer Science}, 410\penalty0 (19):\penalty0
  1826--1846, 2009.

\bibitem[Phuong \& Lampert(2019)Phuong and Lampert]{phuong2019distillation}
Phuong, M. and Lampert, C.~H.
\newblock Distillation-based training for multi-exit architectures.
\newblock In \emph{Proc. International Conference on Computer Vision}, pp.\
  1355--1364, Seoul, South Korea, 2019.

\bibitem[Sener \& Koltun(2018)Sener and Koltun]{sener2018multi}
Sener, O. and Koltun, V.
\newblock Multi-task learning as multi-objective optimization.
\newblock 2018.

\bibitem[Shen et~al.(2021)Shen, Zheng, and Wang]{Shen2021}
Shen, J., Zheng, J., and Wang, X.
\newblock Mmtrans-mt: A framework for multimodal emotion recognition using
  multitask learning.
\newblock In \emph{Proc. International Conference on Advanced Computational
  Intelligence}, pp.\  52--59, Doha, Qatar, 2021.

\bibitem[Wang et~al.(2020)Wang, Tsvetkov, Firat, and Cao]{wang2020gradient}
Wang, Z., Tsvetkov, Y., Firat, O., and Cao, Y.
\newblock Gradient vaccine: Investigating and improving multi-task optimization
  in massively multilingual models.
\newblock \emph{arXiv preprint arXiv:2010.05874}, 2020.

\bibitem[Yu et~al.(2020)Yu, Kumar, Gupta, Levine, Hausman, and
  Finn]{yu2020gradient}
Yu, T., Kumar, S., Gupta, A., Levine, S., Hausman, K., and Finn, C.
\newblock Gradient surgery for multi-task learning.
\newblock \emph{Advances in Neural Information Processing Systems},
  33:\penalty0 5824--5836, 2020.

\bibitem[Zeiler \& Fergus(2014)Zeiler and Fergus]{zeiler2014visualizing}
Zeiler, M.~D. and Fergus, R.
\newblock Visualizing and understanding convolutional networks.
\newblock In \emph{Proc. European conference on computer vision}, pp.\
  818--833. Springer, 2014.

\bibitem[Zhang et~al.(2019)Zhang, Wu, and Schuller]{zhang2019attention}
Zhang, Z., Wu, B., and Schuller, B.
\newblock Attention-augmented end-to-end multi-task learning for emotion
  prediction from speech.
\newblock In \emph{Proc. International Conference on Acoustics, Speech and
  Signal Processing}, pp.\  6705--6709, Brighton, UK, 2019.

\bibitem[Zhu \& Bain(2017)Zhu and Bain]{zhu2017b}
Zhu, X. and Bain, M.
\newblock B-cnn: branch convolutional neural network for hierarchical
  classification.
\newblock \emph{arXiv preprint arXiv:1709.09890}, 2017.

\end{thebibliography}
\bibliographystyle{icml2022}

% \newpage
% \appendix
% \onecolumn

% You can have as much text here as you want. The main body must be at most $8$ pages long.
% For the final version, one more page can be added.
% If you want, you can use an appendix like this one, even using the one-column format.

\end{document}